\documentclass[12pt]{article}
\usepackage{amsfonts}
\usepackage{amssymb}
\usepackage{graphics,amsmath}

\def\hybrid{\topmargin -20pt    \oddsidemargin 0pt
        \headheight 0pt \headsep 0pt
        \textwidth 6.35in       
        \textheight 9.25in       
        \marginparwidth .875in
        \parskip 5pt plus 1pt   \jot = 1.5ex}

\hybrid

\def\baselinestretch{1.2}

\catcode`\@=11

\def\marginnote#1{}
\newcount\hour
\newcount\minute
\newtoks\amorpm
\hour=\time\divide\hour by60
\minute=\time{\multiply\hour by60 \global\advance\minute by-\hour}
\edef\standardtime{{\ifnum\hour<12 \global\amorpm={am}%
        \else\global\amorpm={pm}\advance\hour by-12 \fi
        \ifnum\hour=0 \hour=12 \fi
        \number\hour:\ifnum\minute<10 0\fi\number\minute\the\amorpm}}
\edef\militarytime{\number\hour:\ifnum\minute<10 0\fi\number\minute}

\def\draftlabel#1{{\@bsphack\if@filesw {\let\thepage\relax
   \xdef\@gtempa{\write\@auxout{\string
      \newlabel{#1}{{\@currentlabel}{\thepage}}}}}\@gtempa
   \if@nobreak \ifvmode\nobreak\fi\fi\fi\@esphack}
        \gdef\@eqnlabel{#1}}
\def\@eqnlabel{}
\def\@vacuum{}
\def\draftmarginnote#1{\marginpar{\raggedright\scriptsize\tt#1}}

\def\draft{\oddsidemargin -.5truein
        \def\@oddfoot{\sl preliminary draft \hfil
        \rm\thepage\hfil\sl\today\quad\militarytime}
        \let\@evenfoot\@oddfoot \overfullrule 3pt
        \let\label=\draftlabel
        \let\marginnote=\draftmarginnote
   \def\@eqnnum{(\theequation)\rlap{\kern\marginparsep\tt\@eqnlabel}%
\global\let\@eqnlabel\@vacuum}  }

\def\preprint{\twocolumn\sloppy\flushbottom\parindent 2em
        \leftmargini 2em\leftmarginv .5em\leftmarginvi .5em
        \oddsidemargin -.5in    \evensidemargin -.5in
        \columnsep .4in \footheight 0pt
        \textwidth 10.in        \topmargin  -.4in
        \headheight 12pt \topskip .4in
        \textheight 6.9in \footskip 0pt
        \def\@oddhead{\thepage\hfil\addtocounter{page}{1}\thepage}
        \let\@evenhead\@oddhead \def\@oddfoot{} \def\@evenfoot{} }

\def\numberbysection{\@addtoreset{equation}{section}
        \def\theequation{\thesection.\arabic{equation}}}

\def\underline#1{\relax\ifmmode\@@underline#1\else
        $\@@underline{\hbox{#1}}$\relax\fi}

\def\titlepage{\@restonecolfalse\if@twocolumn\@restonecoltrue\onecolumn
     \else \newpage \fi \thispagestyle{empty}\c@page\z@
        \def\thefootnote{\fnsymbol{footnote}} }

\def\endtitlepage{\if@restonecol\twocolumn \else \newpage \fi
        \def\thefootnote{\arabic{footnote}}
        \setcounter{footnote}{0}}  

\catcode`@=12
\relax

\def\figcap{\section*{Figure Captions\markboth
        {FIGURECAPTIONS}{FIGURECAPTIONS}}\list
        {Figure \arabic{enumi}:\hfill}{\settowidth\labelwidth{Figure
999:}
        \leftmargin\labelwidth
        \advance\leftmargin\labelsep\usecounter{enumi}}}
 \relax
\def\tablecap{\section*{Table Captions\markboth
        {TABLECAPTIONS}{TABLECAPTIONS}}\list
        {Table \arabic{enumi}:\hfill}{\settowidth\labelwidth{Table
999:}
        \leftmargin\labelwidth
        \advance\leftmargin\labelsep\usecounter{enumi}}}
 \relax
\def\reflist{\section*{References\markboth
        {REFLIST}{REFLIST}}\list
        {[\arabic{enumi}]\hfill}{\settowidth\labelwidth{[999]}
        \leftmargin\labelwidth
        \advance\leftmargin\labelsep\usecounter{enumi}}}
 \relax
\makeatletter
\newcounter{pubctr}
\def\publist{\@ifnextchar[{\@publist}{\@@publist}}
\def\@publist[#1]{\list
        {[\arabic{pubctr}]\hfill}{\settowidth\labelwidth{[999]}
        \leftmargin\labelwidth
        \advance\leftmargin\labelsep
        \@nmbrlisttrue\def\@listctr{pubctr}
        \setcounter{pubctr}{#1}\addtocounter{pubctr}{-1}}}
\def\@@publist{\list
        {[\arabic{pubctr}]\hfill}{\settowidth\labelwidth{[999]}
        \leftmargin\labelwidth
        \advance\leftmargin\labelsep
        \@nmbrlisttrue\def\@listctr{pubctr}}}
 \relax
\makeatother
\newskip\humongous \humongous=0pt plus 1000pt minus 1000pt

\newif\ifdtup

\relax

\def\be{\begin{equation}}
\def\ee{\end{equation}}
\def\ba{\begin{eqnarray}}
\def\ea{\end{eqnarray}}

\def\a{\alpha}

\def\no{\noindent}

\def\IR{\relax{\rm I\kern-.18em R}}
\def\II{\relax{\rm 1\kern-.35em1}}

\def\IR{\relax{\rm I\kern-.18em R}}
\def\inv{^{\raise.15ex\hbox{${\scriptscriptstyle -}$}\kern-.05em 1}}

\begin{document}

\begin{titlepage}
\begin{center}

\vskip .5in

{\LARGE Pulsating strings with mixed three-form flux}
\vskip 0.4in

{\bf Rafael Hern\'andez},  \phantom{x} {\bf Juan Miguel Nieto} \phantom{x} and \phantom{x} {\bf Roberto Ruiz} 
\vskip 0.1in

Departamento de F\'{\i}sica Te\'orica I \\
Universidad Complutense de Madrid \\
$28040$ Madrid, Spain \\
{\footnotesize{\tt rafael.hernandez@fis.ucm.es, juanieto@ucm.es, roruiz@ucm.es}}

\end{center}

\vskip .4in

\centerline{\bf Abstract}
\vskip .1in
\no
Circular strings pulsating in $AdS_3 \times S^3 \times T^4$ with mixed R-R and NS-NS three-form fluxes can be described by an integrable deformation of the 
one-dimensional Neumann-Rosochatius mechanical model. In this article we find a general class of pulsating solutions to this integrable system that can be expressed 
in terms of elliptic functions. In the limit of strings moving in $AdS_{3}$ with pure NS-NS three-form flux, where the action reduces to the $SL(2,\mathbb{R})$ WZW model, 
we find agreement with the analysis of the classical solutions of the system performed using spectral flow by Maldacena and Ooguri. 
We use our elliptic solutions in $AdS_{3}$  to extend the dispersion relation beyond the limit of pure NS-NS flux. 
\noindent

\vskip .4in
\noindent

\end{titlepage}

\vfill
\eject

\def\baselinestretch{1.2}


\baselineskip 20pt


The integrable structure that underlies the AdS$_3$/CFT$_2$ correspondence has provided a deep understanding of numerous 
aspects of string theory in backgrounds with an $AdS_3$ factor and two-dimensional conformal field theories with maximal supersymmetry~\cite{BSZ}-\cite{Sfondrini}. 
Integrability has also been shown to remain a symmetry of general string backgrounds that support a mixture of R-R and NS-NS three-form fluxes~\cite{CZ}. 
This discovery has led to an insightful view of many features of the AdS$_3$/CFT$_2$ correspondence in the presence of mixed fluxes~\cite{HT}-\cite{Barik}. 
One of these developments was the demonstration that the sigma-model of type IIB closed strings spinning in $AdS_{3} \times S^{3}$ with mixed fluxes corresponds 
to an integrable deformation of the Neumann-Rosochatius mechanical system~\cite{HN}. The identification of the Lagrangian that describes spinning strings with mixed fluxes 
with a deformation of the Neumann-Rosochatius system allows the use of a systematic approach to the construction of general classes of solutions. In this letter 
we will extend the problem to the case of an ansatz corresponding to closed strings pulsating in $AdS_{3} \times S^{3}$. In particular, we will show that pulsating strings 
with mixed fluxes can also be treated using the integrable deformation of the Neumann-Rosochatius system obtained from the spinning string ansatz. We will make use of the flux-deformation 
of the Uhlenbeck constants of the model to integrate the equations of motion in terms of Jacobi elliptic functions. We will also study the problem in the limit of pure NS-NS three-form flux, 
where the $AdS_{3}$ piece of the system reduces to the $SL(2,\mathbb{R})$ WZW model. On the basis of our class of elliptic solutions we will derive a general form of the dispersion 
relation valid beyond the WZW point. 

In what follows we will first present the Neumann-Rosochatius system that arises from the motion of closed strings pulsating in 
$AdS_3 \times S^3 \times T^4$ with non-vanishing NS-NS flux. We will consider no dynamics along the torus, and thus the background metric will be
\be
ds^2 = - \cosh^2 \rho \, dt^2 + d \rho^2 + \sinh^2 \rho \, d \phi^2 + 
d\theta^2 + \sin^2 \theta d\phi_1^2 + \cos^2 \theta d \phi_2^2 \ , 
\ee
together with 
\be
b_{t \phi} = q \sinh^2 \rho \ , \quad b_{\phi_1 \phi_2} = - q \cos^2 \theta \ , 
\label{Bfield}
\ee
for the NS-NS B-field, where $0 \leq q \leq 1$. The limit $q=0$ corresponds to the case of pure R-R flux, while setting $q=1$ we are left with pure NS-NS flux. 
In the case of pure R-R flux the sigma-model for closed strings rotating in $AdS_3 \times S^3$ reduces to the Neumann-Rosochatius 
system~\cite{NR}. The presence of the NS-NS flux term leads to an integrable deformation of the Neumann-Rosochatius model~\cite{HN}. In this letter we will 
extend the analysis in that reference for the spinning string ansatz to the case of pulsating strings. In order to introduce the pulsating ansatz it will be convenient 
to use the embedding coordinates of $AdS_3$ and $S^3$, 
which are related to the global angles by
\begin{eqnarray}
Y_1 + i Y_2 \! \! \! & = & \! \! \! \sinh \rho \, e^{i \phi} \ , \quad Y_3 + i Y_0 = \cosh \rho \, e^{i  t} \ , \\
X_1 + i X_2 \! \! \! & = & \! \! \! \sin \theta \, e^{i \phi_1} \ , \quad X_3 + i X_4 = \cos \theta \, e^{i \phi_2} \ . 
\end{eqnarray}
In these coordinates the ansatz for a pulsating string is 
\begin{eqnarray}
Y_1 + i Y_2 \! \! \! & = & \! \! \! z_{1}(\tau) \, e^{i \beta_{1}(\tau)+ik_{1} \sigma} \ , \quad Y_3 + i Y_0 = z_{0}(\tau) \, e^{i  \beta_{0}(\tau)} \ , \\
X_1 + i X_2 \! \! \! & = & \! \! \! r_{1}(\tau) \, e^{i \alpha_1(\tau) + i m_{1} \sigma } \ , \quad X_3 + i X_4 = r_{2}(\tau) \, e^{i \alpha_2(\tau) + i m_{2} \sigma} \ , 
\end{eqnarray} 
where we have excluded the winding along the time coordinate because the time direction has to be single-valued. 
When we enter this ansatz in the world-sheet action in the conformal gauge we find 
\ba
L & = & \frac {\sqrt{\lambda}}{2 \pi} \Big[  \sum_{i=1}^2 \frac {1}{2} \big( \dot{r}_i^2 + r_i^2 \dot{\a}_i^2 - r_i^2 m_i^2 \big) 
+ q r_2^2 \, ( m_2 \dot{\alpha}_1 - m_1 \dot{\alpha}_2 ) - \frac {\Lambda}{2} ( r_1^2 + r_2^2 - 1) \nonumber \\
& + & \frac {1}{2} g^{ab} \big( \dot{z}_a \dot{z}_b + z_a z_a \dot{\beta}_{b}^{2}  \big) - z_1^{2} k_1^2
- q k_{1} z_1^2 \dot{\beta}_0 - \frac {\tilde{\Lambda}}{2} \left( g^{ab} z_a z_b +1 \right) \Big] \ ,  
\label{NR}
\ea
where the dot stands for derivatives with respect to $\tau$, the Lagrange multipliers $\Lambda$ and $\tilde{\Lambda}$ are needed to impose that the solutions lie, 
respectively, on $S^{3}$ and $AdS_{3}$, and we have taken $g=\hbox{diag}(-1,1)$, with $a=0,1$. 
We must note that the flux term in~(\ref{NR}) appears with the opposite sign than that in the Lagrangian coming from the spinning string ansatz~\cite{HN}.
The equations of motion for the radial coordinates following from~(\ref{NR}) are given by
\begin{align}
\ddot{r}_1 & = - r_1 m_1^2 + r_1 \dot{\alpha}_1^{2} - \Lambda r_1 \ , \label{r1prime} \\
\ddot{r}_2 & = - r_2 m_2^2 + r_2 \dot{\alpha}_2^{2} - \Lambda r_2 + 2 q r_2 ( m_2 \dot{\alpha}_1 - m_1 \dot{\alpha}_2 ) \ , \label{r2prime}
\end{align}
for the $r_{i}$ coordinates, while for the $z_{a}$ coordinates they are
\begin{align}
\ddot{z}_0 & = z_0 \dot{\beta}^{2}_0 - \tilde{\Lambda} z_0 \ , \\
\ddot{z}_1 & = z_1 \dot{\beta}^{2}_1 - z_1 k_1^2 -\tilde{\Lambda} z_1 -2 q z_1 k_1 \dot{\beta}_0 \ .
\end{align}
The cyclic nature of the spherical and hyperbolic angular coordinates in the Lagrangian implies that its equations of motion can be easily integrated once,
\begin{align}
&\dot{\alpha}_{1}=\frac{v_{1}-q r_{2}^2m_{2}}{r_{1}^2} \ , \quad \dot{\alpha}_{2}=\frac{v_{2}+q r_{2}^2m_{1}}{r_{2}^2} \ , \label{alphaeom} \\
&\dot{\beta}_{0}=-\frac{u_{0}+q k_{1} z^2_{1}}{z_{0}^2} \ , \quad  \dot{\beta}_{1}=\frac{u_{1}}{z_{1}^2} \ , \	\label{betaeom}
\end{align}
where $v_i$ and $u_{a}$ are some integration constants. We can use equations~(\ref{alphaeom}) and~(\ref{betaeom}) to write the energy, the Lorentzian spin and the two angular momenta of the string as  
\be
E = -\sqrt{\lambda}u_{0} \ , \quad S = \sqrt{\lambda}u_{1} \ , \quad J_{1} = \sqrt{\lambda}v_{1} \ , \quad J_{2} = \sqrt{\lambda}v_{2} \ . 
\ee
The equations of motion coming from (\ref{NR}) must be supplied with the Virasoro constraints, which are responsible for the coupling between the $AdS_{3}$ and the $S^{3}$ pieces of the system. 
They take the form
\ba
& \underset{i=1}{\overset{2}{\sum}}\left[\dot{r}_{i}^2+\left(\dot{\alpha}_i^2+m_{i}^2\right)r_{i}^2\right]+\underset{i=1}{\overset{2}{\sum}}g^{ii}(\dot{z}_{i}^2+z_{i}^2\dot{\beta}_{i}^2)+k_{1}^2 z_{1}^2 = 0 \ ,\\
& z_{1}^2 k_{1} \dot{\beta_{1}}+\underset{i=1}{\overset{2}{\sum}}r_{i}^2 m_{i} \dot{\alpha}_{i}=0 \ .
\ea

We will now move to the construction of general solutions to the above system. 
In order to proceed, we will follow~\cite{NR} and introduce ellipsoidal coordinates $\zeta$ and $\mu$ for the sphere and the Anti-de Sitter factors defined, respectively, as the roots of the equations
\be
\frac{r_1^ 2}{\zeta - m_1^ 2} + \frac{r_2^2}{\zeta - m_2^2} = 0 \ , \quad \frac{z_1^ 2}{\mu - k_1^ 2} - \frac{z_0^2}{\mu} = 0 \ .
\ee
The ranges of the ellipsoidal coordinates are $m_1^2 \leq \zeta \leq m_2^2$ and $k_1^2\le \mu$. When we enter directly $\zeta$ and $\mu$ into the equations of motion for 
$r_{i}$ and $z_{a}$ we are 
left with a second order differential equation for each ellipsoidal coordinate. However, we can also use the Uhlenbeck constants of the system to reduce the 
problem to a pair of independent first order differential equations~\cite{NR}. The Uhlenbeck constants for the pulsating Neumann-Rosochatius system 
in the presence of the flux deformation can be obtained immediately from the ones for the spinning Neumann-Rosochatius system in~\cite{HN} just by replacing~$q$ by~$-q$, 
because of the change in the sign of flux terms. Thus the Uhlenbeck constants corresponding to motion of the pulsating string either on the sphere or on the Anti-de Sitter factor are, respectively, given by 
\footnote{The integrability of the Neumann-Rosochatius system follows from the existence of a set of integrals of motion in involution, 
the Uhlenbeck constants \cite{Uhlenbeck}. In the case of strings spinning in $S^3$ there are two integrals $I_1$ and $I_2$, constrained to satisfy $I_1+I_2=1$. 
In the presence of flux they are deformed to $\bar{I}_1$ and $\bar{I}_2$, with the condition $\bar{I}_1+\bar{I}_2=1$~\cite{HN}. 
A similar set of constants, $F_0$ and $F_1$, with the constraint $F_1-F_0=-1$, and their corresponding deformation by the flux term, arises when the string 
spins in~$AdS_{3}$.}
\ba
\bar{I}_1 & \!\! =  \!\! & r_1^2 (1-q^2) +\frac{1}{m_1^2 - m_2^2} \left[ (r_1 \dot{r}_2 - \dot{r}_1 r_2)^2 + \frac{(v_1 - q \omega_2)^2}{r_1^2} r_2^2 + \frac{v_2^2}{r_2^2} r_1^2 \right] \ , \\
\bar{F}_1 & \!\! = \!\! & z_1^2 (1-q^2) +\frac{1}{k_1^2} \left[ (z_1 \dot{z}_0 - \dot{z}_1 z_0)^2 +\frac{(u_0 - q k_1)^2}{z_0^2} z_1^2 +\frac{u_1^2}{z_1^2} z_0^2 \right] \ .
\ea
If we write the Uhlenbeck integrals in terms of the ellipsoidal coordinates we conclude that
\be
\dot{\zeta}^2 = - 4 P_3(\zeta) \ , \quad \dot{\mu}^2 = - 4 Q_3(\zeta) \ ,
\label{diffeqzetamu}
\ee
where the third order polynomials $P_3(\zeta)$ and $Q_3(\zeta)$ are given by
\ba
P_{3}(\zeta) & \!\! =  \!\! & - (1-q^2)(m_2^2-\zeta)(\zeta-m_{1}^2)^2 + \bar{I}_1(m_2^2-m_1^2)(m_2^2-\zeta)(\zeta-m_1^2) \nonumber \\
&  \!\! +  \!\! & (v_1-q m_2)^2(m_2^2-\zeta)^2+v_2^2(\zeta-m_1^2)^2\equiv(1-q^2)\underset{i=1}{\overset{3}{\prod}}(\zeta-\zeta_{i}) \ , \label{P3} \\ 
Q_{3}(\mu) &  \!\! =  \!\! & (1-q^2)(\mu-k_{1}^{2})^2\mu - \bar{F}_1 k_{1}^{2} (\mu - k_{1}^2) \mu + (u_{0}-q k_{1})^2(\mu-k_{1}^{2})^2+u_{1}^2\mu^2 \nonumber \\
&  \!\! \equiv  \!\! & (1-q^2)\underset{i=1}{\overset{3}{\prod}}(\mu-\mu_{i}) \ . \label{Q3}
\ea
The solution to equations (\ref{diffeqzetamu}) can be written in terms of the Jacobian elliptic functions (see~\cite{HN} for details in the case of the spinning string ansatz). We find 
\ba
\zeta(\tau) & \!\! = \!\! & \zeta_{3}+(\zeta_{2}-\zeta_{3}) \, \textnormal{sn}^2\left(\sqrt{\left(1-q^2\right)\left(\zeta_{3}-\zeta_{1}\right)} \, \tau + \tau_{0} , \kappa \right) \ , \label {zetasol} \\
\mu(\tau) & \!\! = \!\! & \mu_{2}+\frac{(\mu_{3}-\mu_{2})(\mu_2-\mu_1)}{\mu_3-\mu_1} \, \textnormal{sd}^2\left(\sqrt{\left(1-q^2\right) \left(\mu_{3}-\mu_{1}\right)} \, \tau + \tau_{0}' , \nu \right) \ ,
\label{musol}
\ea
where $\tau_0$ and $\tau'_0$ are integration constants that can be set to zero by performing a rotation, and the elliptic moduli are 
\begin{equation}
\kappa=\frac{\zeta_{3}-\zeta_{2}}{\zeta_{3}-\zeta_{1}} \ , \quad \nu=\frac{\mu_{3}-\mu_{2}}{\mu_{3}-\mu_{1}} \ .
\label{Elliptic moduli}
\end{equation}
We therefore conclude that~\footnote{Pulsating string solutions in $AdS_{3}\times S^{3}$ with mixed R-R and NS-NS fluxes have been considered before in~\cite{BPS}. 
To contact with the notation in that reference we just need to identify the roots $R_{+}$ and $R_{-}$ in there with our choice of roots through $R_{+}=\mu_{3}/k_{1}^{2} -1$ and $R_{-} = \mu_{1}/k_{1}^{2} - 1$. 
}
\ba
\label{Radii}
r_{1}^2(\tau) & \!\! = \!\! & \frac{\zeta_{3}-m_{1}^2}{m_{2}^2-m_{1}^2}-\frac{\zeta_{3}-\zeta_{2}}{m_{2}^2-m_{1}^2} \, \textnormal{sn}^2\left(\sqrt{\left(1-q^2\right)\left(\zeta_{3}-\zeta_{1}\right)} \, \tau, \kappa \right) \ ,\\
z_{1}^2(\tau) & \!\! = \!\! & \frac{\mu_{2}-k_{1}^2}{k_{1}^2}+\frac{(\mu_{3}-\mu_{2})(\mu_2-\mu_1)}{k_{1}^2(\mu_3-\mu_1)} \, \textnormal{sd}^2\left(\sqrt{\left(1-q^2\right)\left(\mu_{3}-\mu_{1}\right)} \, \tau, \nu \right) \ . 
\ea
We must stress that we need to order the roots of the ellipsoidal coordinate $\zeta$ for the sphere in such a way that $\zeta_1 < \zeta_3$. 
On the contrary, the ellipsoidal coordinate $\mu$ for the Anti-de Sitter factor is unbounded and symmetrical under permutation of the roots $\mu_{1}$ and $\mu_{3}$ 
and thus there is no need to choose the roots $\mu_{1}$ and $\mu_{3}$ according to any particular ordering. Moreover, we should choose $\mu_{2} \ge k_{1}^2$ because $z_{1}^2 \ge 0$. 
Furthermore, we will restrict the elliptic moduli (\ref{Elliptic moduli}) to their fundamental domains $0 \le \kappa, \tau \le 1$, which implies that $\zeta_{1} < \zeta_{2} < \zeta_{3}$, 
and either $\mu_{1} < \mu_{2} < \mu_{3}$ or $\mu_{3} < \mu_{2} < \mu_{1}$.  

We will now focus on the analysis of string solutions restricted to pulsate in $AdS_3 \times S^1$ in the limit of pure NS-NS three-form flux. We will therefore set 
$r_{1}=\alpha_{1}=0$, and $r_{2}=1$ and $\alpha_{2}=\omega$. We must first note that the cubic term in $Q_3(\mu)$ is dressed with a factor $1-q^2$. Thus in the case of pure 
NS-NS flux the degree of the polynomial reduces to two, and the solution can be written in terms of trigonometric functions. In order to understand the reduction of the problem 
from the point of view of the roots of the polynomial, we will first present them for general values of $q$. 
The Virasoro constraints reduce now to 
\begin{eqnarray}
\dot{z}_{0}^2 + \frac{(u_{0} + q k_{1}z_{1}^2)^2} {z_{0}^2} & \!\! = \!\! & \dot{z}_{1}^2+k_{1}^2z_{1}^2+{\omega^2} \ , \label{VirasoroMO} \\ 
u_{1} & \!\! = \!\! & 0 \ \label{u1zero}.
\end{eqnarray}
Using condition~(\ref{u1zero}) it is immediate to check that one of the roots of the polynomial (\ref{Q3}) is given by $\mu_{0} = k_{1}^2$. 
The remaining roots are given by
\be
\mu_{\pm} = \frac {f_1}{2} - \frac{(u_{0}-q k_{1})^2 \mp \sqrt{ \big[(u_{0} - q k_{1})^2 - (1-q^{2}) f_{1} \big]^2  + 4(1-q^2)k_{1}^2(u_{0} - q k_{1})}}{2(1-q^2)} \ ,   
\label{Q3roots}
\ee
where we have defined 
\be
f_{1} = k_{1}^2+\frac{k_{1}^2 \bar{F}_{1}}{1-q^2} \ .
\ee
If we exclude the case with $\bar{F}_{1}=0$, which corresponds to the trivial limit where the solution collapses to a point, it is immediate to check that the roots satisfy $\mu_{-} < \mu_{0} < \mu_{+}$. 
It is also clear that the limit of pure NS-NS flux depends on the sign of the term $(u_0-k_1^2)^2 - k_1^2\bar{F}_1$. If we choose $(u_0-k_1^2)^2>k_1^2\bar{F}_1$, the roots become  
\be
\mu_{-} \rightarrow - \infty \ , \quad \mu_0 \rightarrow k_{1}^2 \ ,\quad \mu_{+} \rightarrow \frac{(u_{0}-k_{1})^2k_{1}^2}{(u_{0}-k_{1})^2-k_{1}^2\bar{F}_{1}} \ .
\ee
In this case, named {\em short string regime} in \cite{MO}, the hyperbolic radius remains bounded, and the solution is given by 
\be
z_{1}^2(\tau)=\frac{k_{1}^2\bar{F}_1}{(u_{0}-k_{1})^2-k_{1}^2\bar{F}_1}\sin^2 \Big( \sqrt{(u_0-k_1^2)^2-k_1^2\bar{F}_1} \, \tau \Big) \ .    
\ee
We can therefore write $z_{1}^2(\tau)=\sinh^2\rho(\tau)=\sinh^2\rho_{0}\sin(\alpha\tau)$, so that
\be
\cosh \rho_0=\frac{|u_0-k_1|}{\sqrt{(u_0-k_1^2)^2-k_1^2\bar{F}_1}} \ .    
\ee
The parameter $\rho_0$ can be interpreted as the maximum hyperbolic radius that the solution can reach. 
On the contrary, in the case where $k_1^2\bar{F}_1>(u_0-k_1^2)^2$ the roots become 
\be
\mu_{-} \rightarrow -\frac{(u_{0}-k_{1})^2k_{1}^2}{k_{1}^2\bar{F}_{1}-(u_{0}-k_{1})^2} \ , \quad \mu_0 \rightarrow k_{1}^2 \ , \quad \mu_{+} \rightarrow \infty \ , 
\ee
and the hyperbolic radius is unbounded. This is the {\em long string regime} of \cite{MO}, and the solution reduces now to 
\be
z_{1}^2(\tau)=\frac{k_{1}^2\bar{F}_1}{k_{1}^2\bar{F}_{1}-(u_{0}-k_{1})^2}\sinh^2 \Big( \sqrt{k_{1}^2\bar{F}_{1}-(u_{0}-k_{1})^2} \, \tau \Big) \ .    
\ee
In this case we will introduce $\rho_{0}$ using $z_{1}^2(\tau) = \sinh^2\rho(\tau) = \cosh^2\rho_{0} \sinh (\alpha\tau)$. Accordingly, 
\be
\sinh \rho_0 = \frac{|u_0-k_1|}{\sqrt{k_1^2\bar{F}_1-(u_0-k_1^2)^2}} \ .    
\ee
Now the parameter $\rho_{0}$ does not have the interpretation of the maximum size of the solution. 
It is also worth to consider the threshold case, with $k_{1}^2\bar{F}_{1}=(u_0-k_1^2)^2$, where the ellipsoidal coordinate $\mu$ 
displays a parabolic behaviour. In this case both $\mu_{-}$ and $\mu_{+}$ diverge and 
\begin{equation}
z_{1}^2(\tau)=k_{1}^2\bar{F}_1\tau^2 \ .    
\end{equation}
In the same way as the short and long string regimes are, respectively, constructed by means of spectral flow on time-like and space-like geodesics, 
the parabolic behaviour can be understood as the result of performing spectral flow on light-like geodesics. 
  
In order to fix the sign of $\alpha$ in the previous expressions we need to find the dependence of the time coordinate $t$ on $\tau$. 
We can obtain that dependence by direct integration of the equation of motion~(\ref{betaeom}) for $\dot{\beta}_{0}$. In the short string regime, 
we conclude that \footnote{We must note that when we move away from the limit of pure NS-NS flux the division of the solutions in three different regimes breaks down. 
However, we can still find $\tan t(\tau)$ in terms of complete elliptic integrals.}
\be
\label{Maldacena-Ooguri}
\tan t(\tau) = \frac{\tan(-k_{1}\tau)+\cosh \rho_{0 }\tan\big(-\textnormal{sign}(u_{0}-k_{1})\sqrt{{(u_{0}-k_{1})^2-\bar{F}_{1}k^2_{1}}}\tau \big)}{1-\cosh \rho_{0} \tan(-k_{1}\tau)
\tan \big( -\textnormal{sign}(u_{0}-k_{1})\sqrt{{(u_{0}-k_{1})^2-\bar{F}_{1}k^2_{1}}}\tau \big)} \ .
\ee
In the long string regime we can proceed identically, or continue analytically expression~(\ref{Maldacena-Ooguri}), to find that 
\be
\tan t(\tau)
=\frac{\tan(-k_{1}\tau)+\sinh\rho_0 \tanh \big( -\textnormal{sign}(u_{0}-k_{1})\sqrt{k_{1}^2\bar{F}_{1}-(u_{0}-k_{1})^2}\tau \big)}{1-\sinh\rho_0\tan(-k_{1}\tau)
\tanh \big( -\textnormal{sign}(u_{0}-k_{1})\sqrt{k_{1}^2\bar{F}_{1}-(u_{0}-k_{1})^2}\tau \big)} \ .
\ee
This is indeed the result in \cite{MO} provided we read $\alpha=-\textnormal{sign}(u_0-k_1)\sqrt{|(u_{0}-k_{1})^2-\bar{F}_{1}k_{1}^2|}$ 
and identify the winding number $k_1$ with $-w$ in that reference. Besides, the threshold case corresponds simply to
\be
t(\tau) = - k_{1} \tau \ .
\ee
  
We will now use our solutions to extend the dispersion relation to general values of the flux parameter, 
beyond the WZW limit of the action. At these point we should emphasize that, despite we have chosen $\bar{F}_1$ and $\omega$ as constants of motion, 
any other set of first integrals is valid as long as they remain independent in phase space. In particular, we can replace $\bar{F}_1$ by $\bar{n}$ in the argument of the elliptic sine, 
$ \hbox{sn} (\bar{n} \tau , \nu) = \hbox{sn} \big( \sqrt{(1-q^{2}) (\mu_{3}-\mu_{1})} \, \tau , \nu \big) $. 
Thus
\be
\bar{n}^4 = \left[k_1^2-\omega^2+2qk_{1}(u_{0}-qk_{1})\right]^2+4(1-q^2)k_{1}^2(u_{0}-qk_{1})^2 \ .
\ee
When we combine this expression with the Virasoro constraint (\ref{VirasoroMO}) and solve for $u_{0}$ we conclude that
\be
E = - \sqrt{\lambda}u_{0}=-\frac{\sqrt{\lambda}}{2} \left[ 2qk_{1}+\frac{1}{k_{1}} \left(-q(k_{1}^2-\omega^2)\mp\sqrt{\bar{n}^4-(1-q^2)(k_{1}^2-\omega^2)} \ \right) \right],
\label{generalE}
\ee
which in the limit of pure NS-NS flux becomes 
\be
E = - \frac{\sqrt{\lambda}}{2}\left(k_{1}+\frac{\omega^2\mp n^2}{k_{1}}\right) \ ,
\label{dispersionrelation}
\ee
where $n$ is just the value of $\bar{n}$ at the WZW point. The upper sign in this expression corresponds to the short string regime, while the lower one is the long string regime, in accordance with~\cite{MO}. 
A natural question arising from this letter is the derivation of (\ref{generalE}) by extending the analysis of the conformal field theory beyond the limit of the WZW model. It would also be interesting 
to investigate in more detail the relation between the  WZW model and the reduction of the Neumann-Rosochatius system with pure NS-NS flux. 


\vspace{8mm}

\centerline{\bf Acknowledgments}

\vspace{2mm}

\no
The work of R.~H. is supported by grant FPA2014-54154-P and by BSCH-UCM through grant GR3/14-A 910770. 


\end{document}